\documentclass[a4paper, 12pt]{article}
\usepackage{}
\usepackage{amsfonts}
\usepackage{dsfont}
\usepackage{mathrsfs}
\usepackage{amssymb}
\usepackage{bbm}
\usepackage{epsfig}
\usepackage{amsmath}
\usepackage{appendix}
\usepackage{color}
\usepackage[english]{babel}

\usepackage{amsfonts}
\usepackage{dsfont}
\usepackage{mathrsfs}
\usepackage{amssymb}
\usepackage{bbm}
\usepackage{epsfig}
\usepackage{amsmath}
\usepackage{appendix}
\usepackage{float}
\usepackage[colorlinks,linkcolor=blue]{hyperref}
\usepackage{fancyhdr}
\newtheorem{them}{Theorem}[section]
\newtheorem{defn}{Definition}[section]
\newtheorem{coro}{Corollary}[section]
\newtheorem{lem}{Lemma}[section]

\newtheorem{rem}{Remark}[section]

\title{\bf  Privacy Preservation in Distributed Subgradient Optimization Algorithms\footnote{This work is supported by grants from the National Science Fund for Distinguished Young Scholars (NSFC No. 71025005), the National Natural Science Foundation of China (NSFC No. 71401163; NSFC No. 71433001), National Program for Support of Top-Notch Young professionals, and the Fundamental Research Funds for the Central Universities in BUCT.}}
\date{}

\author{Youcheng Lou,
     Lean Yu and Shouyang Wang}

\begin{document}

\maketitle

\begin{abstract}
 Privacy preservation is becoming an increasingly
important issue in data mining and machine learning.
In this paper, we consider the privacy preserving features of
distributed subgradient optimization algorithms.
 We first show that
a well-known distributed subgradient synchronous optimization algorithm, in which
all agents make their optimization updates simultaneously at all times,
is not privacy preserving
in the sense that the malicious agent
can learn other agents' subgradients asymptotically. Then we
propose a distributed subgradient projection asynchronous optimization algorithm
without relying on any existing privacy preservation technique,
where agents can exchange data between neighbors directly. In contrast to synchronous algorithms,
in the new asynchronous algorithm agents
make their optimization updates asynchronously. The introduced projection operation and asynchronous optimization mechanism can guarantee that the
proposed asynchronous optimization algorithm is privacy preserving. Moreover, we also establish the optimal convergence
of the newly proposed algorithm. The proposed privacy preservation techniques shed light on developing other privacy preserving distributed optimization algorithms.
\end{abstract}

{\bf Keywords:} Privacy preservation, distributed optimization, asynchronous optimization.

\section{Introduction}

Distributed optimization and learning have attracted much research attention in recent
years due to their wide applications in engineering, machine learning, data mining
and operations research \cite{nedic1,nedic2,joh09,duc,shiw,lou,lou13,yan,li,han}.
In a centralized design, all data collected from the studied problem
needs to be transmitted to a central location. However, this transmission
mechanism may incur prohibitively high cost.
A desirable way is to accomplish the optimization or learning task in a distributed setting, in which
each agent takes partial knowledge about this task and all agents can exchange data with their neighbors
via an underlying network graph.

A widely studied problem is the sum objective optimization problem $\min\sum^n_{i=1}f_i$,
where $f_i$ is agent $i$'s objective function and can be known only
by agent $i$ \cite{nedic1,nedic2,joh09,duc,boyd,shiw,jak,sri,ned}. This group of agents
can solve the optimization problem in a cooperative way by agents'
local optimization updates and local data sharing between neighbors.
Nedic and Ozdaglar proposed a distributed subgradient algorithm with a constant stepsize to solve
this sum objective optimization problem and presented a convergence error
between the generated estimates and the optimal objective value in terms of
model parameters in \cite{nedic1}. Then Nedic et. al proposed a distributed
subgradient time-varying stepsize algorithm
to solve a more general sum objective constrained optimization problem in \cite{nedic2}. The optimal convergence was established
under mild assumptions of boundedness of subgradients, joint connectivity of network graphs and the classical stochastic approximation
conditions. Duchi et al. proposed a dual averaging algorithm, where various
sharp convergence bounds as a function of the network size
and network graphs were provided in \cite{duc}. Moreover, distributed alternating direction method of multipliers
were also studied with faster convergence rate compared to gradient-based algorithms in \cite{boyd,shiw}.

In these existing distributed optimization algorithms \cite{nedic1,nedic2,joh09,duc,boyd,shiw},
in order to accomplish the task agents need to share their data
with their neighbors. However, this may lead to privacy disclosure.
 Privacy preservation is becoming an increasingly
important issue in applications involving sensitive data, especially in distributed settings \cite{agr,vai}.
Clearly, it is desirable that on one hand, agents can jointly solve the
optimization problem, while on the other hand, agents' privacy
can be effectively preserved. In fact, some work have also been done
on designing privacy preserving algorithms to solve optimization problems \cite{yan,huang,wee,li,han,hans}.

The existing privacy preserving methods can be roughly classified into two classes:
cryptograph-based approaches \cite{glo} and non-cryptograph-based approaches \cite{huang,li,hans,man,dwo}.
 In cryptograph-based methods, many mechanisms are designed to encrypt the
data needed to be transmitted and decrypt the received data so that the privacy
is not disclosed. The low efficiency of cryptograph-based methods has motivated
much research on developing non-cryptograph-based methods. An important non-cryptograph-based
method is the $\varepsilon$-differential privacy approach \cite{huang,li,hans,dwo}, which typically employs a randomization perturbation method. An disadvantage of this approach is that the sensitivity of the considered
algorithm is usually hard to accurately estimate, and then in order to
ensure the pre-specified privacy preserving level quantified by $\varepsilon$, the added random noise is required
to have higher covariance, which in return degrades the optimality significantly \cite{wee}.

In this paper, we will consider the privacy preserving features
of distributed subgradient optimization algorithms.
Agents' privacy
may refer to the parameters in the objective functions \cite{wee},
convex constraint sets \cite{hans} or the subgradients of objective functions \cite{yan}.
Similar to \cite{yan}, in this paper the subgradients of agents' objective functions are defined as agents' privacy that needs to be protected.
In our problem domain, we assume that there is a macilious agent that does not follow
the algorithm truthfully and can transmit any data to its neighbors.
This malicious agent will keep a record of all data shared with its
neighbors in order to discover other agents' subgradients.

We will first show that in the well-known
distributed subgradient synchronous optimization algorithm in which all agents make their optimization updates simultaneously,
 the malicious agent can asymptotically discover
other agents' subgradients for almost all adjacency matrices when this malicious
agent can communicate with all other agents and the stepsize is diminishing.
In this sense this synchronous optimization algorithm is not privacy preserving.
Then we will design
a new distributed subgradient projection asynchronous optimization algorithm
and establish its optimal convergence. Different from the existing
synchronous algorithms where all agents make their optimization updates
simultaneously after taking a weighted average of the received data from their neighbors, in our asynchronous algorithm at each time all agents make their optimization updates asynchronously and the optimization update time sequences are different
for different agents. When currently some agent does not
make its own optimization update, this agent just takes the weighted average of the received data from its neighbors
as the next step's estimate without any optimization update.
Moreover, we also artificially introduce a projection set,
which combines with the asynchronous optimization update mechanism
can effectively prevent the privacy disclosure.
The main contribution of this paper is that
we show that the well-known distributed subgradient synchronous optimization algorithm
is not privacy preserving under some cases, and propose a new privacy preserving
distributed subgradient projection asynchronous optimization algorithm without employing any cryptograph-based and differential privacy
technique following detailed convergence analysis.

 Our work is closely related to the recent work \cite{yan}, in which Yan et al. considered
the privacy preservation problems of their proposed distributed subgradient online learning
synchronous optimization algorithm and showed that their algorithm has intrinsic privacy-preserving properties.
The authors also presented the necessary and sufficient conditions to ensure
the privacy preserving properties. Different from the work by Yan et al. \cite{yan}, we consider the static distributed
optimization instead of dynamical (online learning) optimization in order to highlight the main contribution.
In fact, the current results can be generalized to the dynamical cases.
In this paper, we relax the assumption that the malicious agent knows the adjacency matrix
of the network graph used in \cite{yan} considering that in practice any agent is hard to obtain this adjacency matrix, especially in large scale networks and distributed settings. Compared with \cite{nedic1,nedic2,joh09},
besides the optimal convergence, we also consider the privacy preserving properties
of distributed algorithms. Moreover, different from the cryptograph-based and differential privacy
techniques studied in \cite{huang,li,hans,man,dwo,glo}, the data in our algorithm can be transmitted
directly between neighbors and we do not use any additional
privacy preservation technique to disguise agents' data.

The rest of the paper is organized as follows. In Section 2,
we present some preliminaries on the well-known distributed subgradient synchronous optimization algorithm (DSSOA)
and the problem formulation of the interested privacy preservation problems.
In Subsections 3.1 and 3.2, we consider the adjacency matrix discovery problems
of a special case of DSSOA, i.e., the distributed consensus algorithms, and the DSSOA, respectively.
In Section 4, we present our privacy preserving distributed subgradient projection asynchronous optimization algorithm
and establish its optimal convergence. Finally,
some concluding remarks are given in Section 5.

\section{Preliminaries and Problem Formulation}

In this section, we first introduce a well-known distributed subgradient synchronous optimization algorithm and
then state the interested privacy preserving problems of this algorithm.

\subsection{A Distributed Subgradient Synchronous Optimization Algorithm}

Consider a network consisting of $n$ agents with node set $\mathcal{V}=\{1,...,n\}$.
The communication among agents can be described by a directed graph $\mathcal{G}=(\mathcal{V},\mathcal{E})$,
 where arc $(j,i)\in\mathcal{E}$ means that agent $i$ can receive the data sent by agent $j$.
Here node $j$ is said to be node $i$'s neighbor if $(j,i)\in\mathcal{E}$.
Let $\mathcal{N}_i=\{j|(j,i)\in \mathcal{E}\}$ denote the set of node $i$'s neighbors.
Associated with graph $\mathcal{G}$, there is usually a nonnegative adjacency matrix $\bar A=(a_{ij})\in \mathbb{R}^{n\times n}$ to characterize the weights among agents, where the entries $a_{ij}$ are nonnegative and $a_{ij}$ is positive if and only if $(j,i)\in \mathcal{E}$. Graph $\mathcal{G}$ is said to be strongly
connected if there exists a path from $i$ to $j$ for each pair of
nodes $i,j\in \mathcal{V}$. The objective of this network is to cooperatively solve
the sum optimization problem
\begin{align}\label{opti}
\min_{x\in \mathbb{R}^m}\sum^n_{i=1}f_i(x)
\end{align}
 where $f_i:\mathbb{R}^m\rightarrow \mathbb{R}$
is the convex objective function of agent $i$ to be minimized.
In a distributed setting, each agent only knows its own objective function.

A possible algorithm for solving
 (\ref{opti}) is the following distributed subgradient synchronous optimization algorithm (DSSOA) \cite{nedic1}:
\begin{align}\label{eqq1}
x_i(k+1)&=\sum_{j\in \mathcal{N}_i}a_{ij}x_j(k)-\alpha_k d_i(k),\;k\geq0,\;d_i(k)\in \partial f_i(x_i(k)),\;\;i=1,...,n,
\end{align}
where $x_i(k)$ is agent $i$'s estimate for the optimal solution of (\ref{opti}) at time $k$; $0<\alpha_k\leq\alpha^*$ is the stepsize, $\alpha^*>0$;
$\partial f_i(x_i(k))$ is the subdifferential that contains all subgradients of $f_i$
at $x_i(k)$. In algorithm (\ref{eqq1}), before agents generate their estimates at the next step,
they first take a weighted average of the estimates received from their neighbors,
and then make an optimization update following a negative gradient direction.
Here we say that algorithm (\ref{eqq1}) is synchronous since all agents make their optimization
updates simultaneously.

\begin{rem}
In \cite{nedic1},  Nedic and A. Ozdaglar proposed the algorithm (\ref{eqq1})
with a constant stepsize $\alpha_k\equiv\alpha$ to solve optimization problem (\ref{opti}),
where the convergence error between agents' estimates and the optimal function value
is presented in terms of the constant stepsize and some other algorithm parameters.
In \cite{nedic2}, Nedic et al. considered a more general constrained optimization
problem $\min_{x\in K}\sum^n_{i=1}f_i(x)$
and proposed a distributed subgradient projection algorithm
$$x_i(k+1)=P_K\Big(\sum_{j\in \mathcal{N}_i}a_{ij}x_j(k)-\alpha_k d_i(k)\Big),k\geq0$$
with a time-varying stepsize.
Following this, many distributed subgradient algorithms emerge under various scenarios,
for example, to deal with inexact subgradients with errors \cite{ram} and random network graphs \cite{lob,sri}.
\end{rem}

We next introduce three basic assumptions on the connectivity of the network graph, adjacency matrix and the boundedness of subgradients
\cite{nedic1,nedic2,joh09,lou,yan}.

\noindent {\bf Assumption 1}: The graph $\mathcal{G}$ is strongly connected.

\noindent {\bf Assumption 2}:
The adjacency matrix $\bar A$ is doubly stochastic,
i.e., $\sum^n_{j=1}a_{ij}=\sum^n_{j=1}a_{ji}=1$
for all $i$.

\noindent {\bf Assumption 3}: The subgradients of $f_i$ are bounded, i.e., there is
$L>0$ such that
$$
\sup_{q\in\bigcup_i\partial f_i(x)}|q|\leq L, \;\;\forall x\in \mathbb{R}^m,
$$
here we use $|\cdot|$ to denote the Euclidean norm of a vector in $\mathbb{R}^m$.

Although each agent only knows its own objective function, surprisingly this simple weighted average information exchange
mechanism can make the network achieve an optimal consensus, as indicated in the following theorem,
which can be found in Proposition 2 in \cite{nedic2}.
\begin{them}\label{optimal}
Consider DSSOA (\ref{eqq1}) with Assumptions 1, 2, 3, $\sum^\infty_{k=0}\alpha_k=\infty$ and
$\sum^\infty_{k=0}\alpha_k^2<\infty$. Then the network will achieve an optimal consensus, i.e.,
there exists $x^*\in \arg\min\sum^n_{i=1}f_i$
such that $\lim_{k\rightarrow\infty}x_i(k)=x^*$,
$i=1,...,n$.
\end{them}

\subsection{Problem Formulation}

In algorithm (\ref{eqq1}), agents need to share their data (the estimates for the optimal solutions) with their neighbors in order to solve the sum objective optimization problem. This direct information exchange
may lead to privacy leakage. Recently, privacy preservation becomes an increasingly important issue in
machine learning and data mining fields in distributed settings.

It is desirable that on one hand, agents can jointly accomplish the desired task,
while on the other hand, agents' private information can be effectively protected.
However, most of the existing distributed optimizations algorithms including the
subgradient algorithm (\ref{eqq1}) mainly focus on the algorithm design and optimal
convergence analysis, but not the privacy preservation problems (referring to those algorithms in \cite{nedic1,nedic2,joh09,duc})
except the $\varepsilon$-differentially
private methods, in which typically a random perturbation technique
is used to prevent privacy disclosure \cite{huang,li,hans}
and cryptograph-based methods \cite{glo}.
A disadvantage of differentially
private methods is that the sensitivity of the studied
algorithm is extremely hard to accurately estimate, and consequently, it is necessary that the added noise has higher covariance for ensuring the desired privacy preservation level. This in return degrades the optimality significantly. Moreover, it is usually a trade-off between the desired privacy accuracy requirement specified by the parameter $\varepsilon$
and the optimality of the solutions.
Moreover, a main disadvantage of cryptograph-based methods
is that agents need to encrypt the data needed to be shared and decrypt the received data
from their neighbors frequently, which incurs in low efficiency.

In this paper, we define
agents' subgradients as their private information, similar to the setting in \cite{yan}.
In our problem domain, we assume there is a malicious agent that may not follow
the algorithm truthfully and can transmit any data to its neighbors.
We call those agents that follows the algorithm truthfully as regular agents.
The malicious agent will keep a record of all the exchanged data with its neighbors
and try to discover its neighbors' subgradients.

In this paper, we are interested in the following two problems:

(i) Is the existing
distributed synchronous algorithm (\ref{eqq1}) privacy preserving in the sense that
the malicious agent can discover other agents' subgradients based on the exchanged data between neighbors?

(ii) If algorithm (\ref{eqq1}) is not privacy preserving,
can we design a privacy preserving distributed subgradient algorithm that does not
employ any cryptograph-based and differentially private technique?

For the first problem, in Section 3 we will investigate the privacy preserving features of the
DSSOA (\ref{eqq1}), in which information are shared
directly between neighbors and all agents make their optimization updates simultaneously.
The results show that for almost all adjacency matrices the malicious agent can learn the adjacency matrix of the
network graph asymptotically when this malicious agent can receive all other agents' estimates.
This implies that DSSOA (\ref{eqq1})
is not privacy preserving since this malicious agent can discover
other agents' subgradients by some simple calculations.

For the second problem, in Section 4 we will propose a new distributed subgradient projection asynchronous optimization algorithm,
in which agents first take a weighted average of the received estimates from their neighbors
and then either make an optimization update following a subgradient direction to generate the next step's estimates
or just take the weighted average
as the next step's estimates.
Different from DSSOA (\ref{eqq1}), in the newly proposed asynchronous optimization algorithm,
agents make their optimization updates asynchronously and
the optimization time sequences for different agents may be different.
This newly proposed asynchronous optimization mechanism and the introduced projection set
can effectively protect agents' subgradient information.

\section{Adjacency Matrix Discovery}

In this section, we investigate the adjacency matrix
discovery problem of DSSOA (\ref{eqq1}). Clearly, if the malicious agent can obtain the adjacency matrix
 of the network graph and observe all other regular agents' estimates,
then the malicious node can discover other agents' subgradients by simple subtraction
calculations by noticing that the stepsizes for all agents are the same in synchronous algorithm (\ref{eqq1}).
We will first consider the adjacency matrix
discovery problem of distributed consensus algorithms, which is a special case
of DSSOA (\ref{eqq1}) with trivial objective functions, and establish some necessary and sufficient
conditions to ensure that the adjacency matrix can/cannot be discovered. Then for DSSOA (\ref{eqq1}) we will show that the malicious agent
can discover the adjacency matrix asymptotically under mild conditions by transmitting some appropriate data sequence to other regular agents.

In the work by Yan et al. \cite{yan}, it is assumed that the malicious agent knows the
adjacency matrix in advance. Different from it, here we do not enforce this assumption since in
practice, usually it is extremely hard to obtain
this adjacency matrix, especially in large scale directed networks, taking into account the following two reasons:
first, the adjacency matrix captures the global network information and
then generally agents cannot obtain it in a distributed setting; second, agents are not willing to
leak the weights assigned to their neighbors to other agents from the viewpoint of privacy preservation.

In this section, we without loss of generality assume that agent $n$
is the malicious agent, agents $1,2,...,n-1$ (regular agents) are the malicious agent's neighbors
and
the induced subgraph generated by all regular agents is strongly connected.

\subsection{A Special Case: Distributed Consensus Algorithms}

In this subsection, we first consider a special case of DSSOA (\ref{eqq1})
with trivial (constant) objective functions. In this case, algorithm (\ref{eqq1}) induces to
\begin{align}\label{eq1}
x_i(k+1)=\sum_{j\in \mathcal{N}_i}a_{ij}x_j(k),\;i=1,...,n,\;k\geq0.
\end{align}

Without loss of generality, in this subsection we assume $m=1$ for notational simplicity.
Next we will investigate whether the malicious agent $n$ can discover the
adjacency matrix based on the exchanged data with other agents.
Note that the malicious agent does not follow
the algorithm truthfully and can transmit any data to all other regular agents
with the aim to discover other agents' subgradient information.
Let $\{u(k)\}_{k\geq0}$ be a data sequence that the malicious agent $n$ transmits to other agents
(i.e., $x_n(k)=u(k)$ for all $k\geq0$).
Partition adjacency matrix $\bar A$ into
\begin{align}
&\bar A=\begin{pmatrix}
           A & b \\
           * & * \\
\end{pmatrix},\nonumber\\
&b=(a_{1n},...,a_{(n-1)n})'\in \mathbb{R}^{n-1},\;\;A\in \mathbb{R}^{(n-1)\times(n-1)},\nonumber
\end{align}
where $'$ denotes the transpose of a vector.
Then we rewrite (\ref{eq1}) as a compact form:
\begin{align}\label{con}
x(k+1)=Ax(k)+bu(k),k\geq0,
\end{align}
where $x(k)=(x_1(k),...,x_{n-1}(k))'$. We also denote $b=(b_1,...,b_{n-1})'$ for simplicity. In control community, (\ref{con})
is referred to as a single-input control system.
Note that $b_i>0$ for all $i$ since we assume in this section that
all regular agents are the malicious agent's neighbors.
When there is no confusion, here we roughly call the weight pair $(A,b)$ describing the weights
within regular agents and that between regular agents and the malicious agent
as the adjacency matrix.
In the following, we formally introduce the definition of adjacency matrix discovery.
Recall that a vector is said to be a stochastic vector if it is nonnegative and
the sum of its components is one, and a matrix is said to be a stochastic matrix
if all its rows are stochastic vectors.

\begin{defn}(\textbf{Adjacency Matrix Discovery})
We say that the adjacency matrix $(A,b)$ of (\ref{con}) cannot be discovered by the malicious agent
if there exists another stochastic matrix $(A^*,b^*)\neq(A,b)$ with each component of $b^*$ being positive such that for any sequence $\{u(k)\}_{k\geq0}$,
$x^*(k)=x(k)$ for all $k\geq0$,
where $\{x^*(k)\}_{k\geq0}$ are the estimates generated by
the algorithm
$$x^*(k+1)=A^*x^*(k)+b^*u(k),k\geq0
$$
 with $x(0)=x^*(0)$, and can be discovered by the malicious agent otherwise.
\end{defn}

\begin{them}\label{th}
The adjacency matrix $(A,b)$ of algorithm (\ref{con}) cannot be discovered if and only if the following matrix equations
 with variable $z$ have at least two solutions:
$$
\left\{
  \begin{array}{ll}
    (A-z)A^kb=0, & \hbox{$k=0,1,...,n-2,$} \\
    (A-z)A^kx(0)=0, & \hbox{$k=0,1,...,n-2,$}
  \end{array}
\right.
$$
$$
\mbox{subject\;to}\; z\in \mathbb{R}^{(n-1)\times(n-1)}, (z,b)\;\mbox{is a stochastic matrix}.
$$
\end{them}

\emph{Proof.}
(Necessity). According to the definition of adjacency matrix discovery,
there exists another stochastic matrix $(A^*,b^*)\neq(A,b)$ such that the two estimate sequences generated by algorithm
(\ref{con}) with respective $(A,b)$ and $(A^*,b^*)$ are identical for any sequence $\{u(k)\}_{k\geq0}$.
Then
$$
(A-A^*)x(k)+(b-b^*)u(k)=0,k\geq0.
$$
As a result, $b=b^*$, and consequently, $(A-A^*)x(k)=0$ for any $k\geq0$.
Therefore, $(A-A^*)x(0)=0$. From $(A-A^*)x(1)=0$ and $x(1)=Ax(0)+bu(0)$,
we can find that $(A-A^*)b=0$ and $(A-A^*)Ax(0)=0$.
Analogously, from $(A-A^*)x(2)=(A-A^*)(A^2x(0)+Abu(0)+bu(1))$
we can obtain that $(A-A^*)Ab=0$, $(A-A^*)A^2x(0)=0$.
Other equations can be obtained in a similar way.

(Sufficiency). The sufficiency can be shown directly from the sufficiency condition and the fact that each $A^k,k\geq n-1$
can be expressed as a linear combination of $A^r,r=0,1,...,n-2$.
We complete the proof. \hfill$\square$

Let span$\{p_1,...,p_\ell\}$ and rank$\{p_1,...,p_\ell\}$ denote the subspace generated by vectors $p_1,...,p_\ell$,
and the rank of vectors $p_1,...,p_\ell$, respectively.
Also let $\textbf{1}$ denote the vector of all ones in $\mathbb{R}^{n-1}$.
The following two corollaries can be obtained directly
from Theorem \ref{th}.

\begin{coro}\label{co1}
If \begin{align}\label{rank}
&\mbox{span}\big\{\textbf{1},b,Ab,...,A^{n-2}b,x(0),Ax(0),...,A^{n-2}x(0)\big\}=\mathbb{R}^{n-1},
\end{align}
then the adjacency matrix $(A,b)$ of algorithm (\ref{con}) can be discovered.
\end{coro}

\begin{coro}\label{co2}
If single-input control system (\ref{con}) is completely controllable
(equivalently, rank$(b,Ab,...,A^{n-2}b)=n-1$),
then the adjacency matrix $(A,b)$ of algorithm (\ref{con}) can be discovered.
\end{coro}

From Corollaries \ref{co1} and \ref{co2} we can find that for almost all adjacency matrices
except a zero Lebesgue measure weight set, the adjacency matrix $(A,b)$ of algorithm (\ref{con}) can be discovered by the malicious agent.
The following is a necessary and sufficient condition
that the adjacency matrix can be discovered for a special class of graphs.

\begin{them}\label{th4}
Assume there is a node $i,i\neq n$ in graph $\mathcal{G}$ such that each node $j,j\neq i,j\neq n$
is a neighbor of this node. Then the adjacency matrix $(A,b)$ of algorithm (\ref{con})
can be discovered if and only if (\ref{rank}) holds.
\end{them}
\emph{Proof.}
The sufficiency can be obtained from Corollary \ref{co1}. We now show by contradiction the necessity.
We assume without loss of generality that nodes $2,...,n-1$ are node 1's neighbors.
As a result, all components of the first row of $A$, which is denoted as $\textbf{a}$, are positive.
Select a nonzero vector
$\textbf{c}\in \mbox{span}\{\textbf{1},b,Ab,...,A^{n-2}b,x(0),Ax(0),...,A^{n-2}x(0)\}^{\perp}$
with sufficiently small components such that all components of $\textbf{a}-\textbf{c}$
are positive ($^\bot$ denotes the orthogonal complement of a subspace).
Then the matrix $z$ with the first row being $\textbf{a}-\textbf{c}$ and all other rows are the same as that of $A$
is also a solution of the matrix equations in Theorem \ref{th}. This contradicts Theorem \ref{th} and then the
necessity follows.
The proof is completed. \hfill$\square$

We next present a necessary and sufficient condition when the network contains three agents.

\begin{them}\label{th3}
Consider algorithm (\ref{con}) with a completely connected graph and $n=3$.
Then the adjacency matrix $(A,b)$ of algorithm (\ref{con})
cannot be discovered if and only if $b_1=b_2$,
$x_1(0)=x_2(0)$ and $a_{11}+a_{12}=a_{21}+a_{22}$.
\end{them}
\emph{Proof.}
The sufficiency is straightforward. In fact, when the sufficient conditions hold,
any nonnegative matrix
$\begin{pmatrix} z_1 & z_2 \\ z_3 &  z_4 \end{pmatrix}$
 satisfying $z_1+z_2=z_3+z_4=a_{11}+a_{12}$
is a solution of the matrix equations in Theorem \ref{th}.
Now we show the necessity by contradiction.
Hence first suppose $x_1(0)\neq x_2(0)$.
Then from $(A-z)x(0)=0$ and $(z,b)$ is a stochastic matrix,
we have that $a_{11}x_1(0)+a_{12}x_2(0)=z_1x_1(0)+z_2x_2(0)$ and $a_{11}+a_{12}=z_1+z_2$. That is,
$$
\begin{pmatrix}
    x_1(0) & x_2(0) \\
    1 & 1 \\
\end{pmatrix}
\begin{pmatrix}
           a_{11}-z_1 \\ a_{12}-z_2\\
\end{pmatrix}=0.
$$
The above equation implies that $z_1=a_{11}$, $z_2=a_{12}$ due to $x_1(0)\neq x_2(0)$.
Similarly, we can show $z_3=a_{21}$, $z_4=a_{22}$. This implies that the matrix equations in Theorem \ref{th} has a unique solution,
which yields a contradiction. Then $x_1(0)=x_2(0)$. Analogously, from $(A-z)Ax(0)=0$ we can also prove that
the two entries of $Ax(0)$ are the same. Therefore, it follows that $a_{11}+a_{12}=a_{21}+a_{22}$.
From the first matrix equation in Theorem \ref{th} we can also show that $b_1=b_2$ in a similar way.
The proof is completed. \hfill$\square$

We now consider the problem the malicious agent how to discover the adjacency matrix $(A,b)$
by choosing an appropriate sequence $\{u(k)\}_{k\geq0}$
when condition (\ref{rank}) holds.

\begin{them}\label{cd}
Assume (\ref{rank}) holds. Then the adjacency matrix $(A,b)$ of distributed algorithm (\ref{con}) can be discovered by choosing
\begin{align}
&u(0)=u(1)=\cdots=u(n-2)=0,\nonumber\\
&u(n-1)=\cdots=u(2n-2)=1.\nonumber
\end{align}

\end{them}
\emph{Proof.}
From Corollary \ref{co1} we know that the adjacency matrix $(A,b)$ of algorithm (\ref{con}) can be discovered
when (\ref{rank}) holds.
Clearly, by (\ref{con}) we have the matrix equation
\begin{align}
&\big(\textbf{1},x(1),...,x(n),x(n+1),...,x(2n-1)\big)\nonumber\\
&=\big(A,b\big)\begin{pmatrix}
                                              \textbf{1}&x(0) & \cdots & x(n-1)& x(n)& \cdots& x(2n-2) \\
                                              1&u(0) & \cdots & u(n-1)& u(n)&\cdots & u(2n-2) \\
\end{pmatrix}\nonumber\\
&=
\begin{tiny}\big(A,b\big)\begin{pmatrix}
                                              \textbf{1}&x(0) &\cdots & A^{n-1}x(0)
                                              +\sum^{n-2}_{r=0} A^{n-2-r}bu(r) &A^nx(0)+\sum^{n-1}_{r=0} A^{n-1-r}bu(r)&\cdots &A^{2n-2}x(0)+\sum^{2n-3}_{r=0} A^{2n-3-r}bu(r)\\
                                              1&u(0) &\cdots & u(n-1) &u(n)&\cdots & u(2n-2)\\
\end{pmatrix}\end{tiny}\nonumber
\end{align}
Rewrite the above matrix equation as $Z=\big(A,b\big)Y$. If matrix $Y$ has full row rank, then $\big(A,b\big)$
is uniquely determined by
$$
(A,b)=ZY'(YY')^{-1}.
$$
Note that the square matrix $YY'$ is invertible if and only if $Y$ is full row rank.
We next show that the matrix $Y$ has full row rank under condition (\ref{rank}) by choosing $u(0),...,u(2n-2)$ given in this theorem.

By letting $u(0)=u(1)=\cdots=u(n-2)=0$ and $u(n-1)=\cdots=u(2n-2)=1$ in $Y$ yields
the matrix
\begin{tiny}
\begin{align}\label{da}
\begin{pmatrix}
      \textbf{1} & x(0) & Ax(0)& \cdots & A^{n-2}x(0) & A^{n-1}x(0) & A^nx(0)+b & \cdots &
                                              A^{2n-2}x(0)+\sum^{2n-3}_{r=n-1} A^{2n-3-r}b \\
      1 & 0 & 0 & \cdots& 0 & 1 & 1 & \cdots & 1 \\
\end{pmatrix}
\end{align}
\end{tiny}
Noticing that any $A^k$, $k\geq n-1$ can be expressed as a linear combination of $I$ (the identity matrix), $A,...,A^{n-2}$, we can find that the matrix in (\ref{da}) is certainly full row rank.
We complete the proof. \hfill$\square$

\subsection{Adjacency Matrix Discovery of DSSOA}

In this subsection, we consider the adjacency matrix discovery problem of
DSSOA (\ref{eqq1}).
Clearly, (\ref{eqq1}) can be written as the following compact form:
\begin{align}\label{eqq2}
x(k+1)=Ax(k)+bu(k)-\varepsilon(k),k\geq0,
\end{align}
where $\varepsilon(k)=(\varepsilon_1(k),...,\varepsilon_{n-1}(k))'$,
$\varepsilon_i(k)=\alpha_k d_i(k)$.

We first present a useful lemma for the following analysis.

\begin{lem}\label{bou}
Assume Assumptions 1, 2, 3 hold.
Then the estimates $x_i(k),i,k$ generated by algorithm (\ref{eqq2}) are bounded if the $u(k),k\geq0$ transmitted by the malicious agent
 to other agents are bounded.
\end{lem}
\emph{Proof.} By Assumption 3, we have $|\varepsilon_i(k)|\leq \alpha_kL\leq\alpha^*L$.
It is also easy to see $||A||_\infty:=\max_{1\leq i\leq n-1}\sum^{n-1}_{j=1}a_{ij}=\max_{1\leq i\leq n-1}(1-b_i)<1$.
From (\ref{eqq2}) by induction we have $x(k+1)=A^{k+1}x(0)+\sum^{k}_{r=0} A^{k-r}(bu(r)-\varepsilon(r))$, $k\geq0$.
The proceeding three relations imply that for any $k$,
\begin{align}
||x(k+1)||_\infty
&\leq ||A||_\infty^{k+1}||x(0)||_{\infty}
+\sum^{k}_{r=0}||A||_\infty^{k-r}\big(u^*+\alpha^*L\big)\nonumber\\
&\leq ||x(0)||_{\infty}
+\frac{u^*+\alpha^*L}{1-||A||_\infty}\nonumber\\
&<\infty,\nonumber
\end{align}
where $u^*:=\sup_{k\geq0}|u(k)|$ is a finite number by hypothesis. Then the proof is completed.\hfill$\square$

\begin{them}\label{th2}
Consider distributed algorithm (\ref{eqq2}) with Assumptions 1, 2, 3. Suppose rank$(b,Ab,...,A^{n-2}b)=n-1$ and $\lim_{k\rightarrow\infty}\alpha_k=0$. Then the adjacency matrix $(A,b)$ of algorithm (\ref{eqq2}) can be discovered asymptotically by choosing an appropriate sequence $\{u(k)\}_{k\geq0}$.
\end{them}
\emph{Proof.}
Denote $s_{r,k}=r(2n-1)+k,r\geq0,k=0,...,2n-2$
and let $u(s_{r,0})=u(s_{r,1})=\cdots=u(s_{r,n-2})=0$,
$u(s_{r,n-1})=u(s_{r,n})=\cdots=u(s_{r,2n-2})=1$ for each $r\geq0$.
Similar to the analysis in the proof of Theorem \ref{cd},
\begin{align}\label{ad}
&\big(A,b\big)=Z_rY'_r(Y_rY'_r)^{-1}+\big(0,\varepsilon(s_{r,1}-1),...,\varepsilon(s_{r,2n-1}-1)\big)Y'_r(Y_rY'_r)^{-1},
\end{align}
where $Z_r=\big(\textbf{1},x(s_{r,1}),x(s_{r,2}),...,x(s_{r,2n-1})\big)$, $Y_r$ has the same definition as the matrix given in (\ref{da})
by replacing $x(0)$ with $x(s_{r,1}-1)$.
We can find that if rank$(b,Ab,...,A^{n-2}b)=n-1$, then $Y_rY'_r$ is full row rank and hence the
inverse $(Y_rY'_r)^{-1}$ exists.

By Lemma \ref{bou}, the estimates $x_i(k),i,k$ are bounded. Then $Y_rY'_r,r\geq0$ are bounded and as a result,
we can show by contradiction that $(Y_rY'_r)^{-1},r\geq0$ are also bounded based on the following two conclusions:

(i)
If $B_rC_r=I$ (the identity matrix) for any $r$ and $\lim_{r\rightarrow\infty}B_r=B$, where the inverse $B^{-1}$ exists, then
$\lim_{r\rightarrow\infty}C_r=B^{-1}$;

(ii) Under the rank condition rank$(b,Ab,...,A^{n-2}b)=n-1$,
the inverse of $Y_rY'_r$ exists for any $x(s_{r,1}-1)$ when we view $Y_rY'_r$
as a matrix function with variable $x(s_{r,1}-1)$ in a bounded closed set.

The boundedness of $(Y_rY'_r)^{-1},r\geq0$ combines with the hypothesis condition $\lim_{k\rightarrow\infty}\alpha_k=0$
 imply that the second term in (\ref{ad}) tends to zero.
Then we conclude that
$(A,b)$ can be discovered asymptotically in the sense that $\lim_{r\rightarrow\infty}|Z_rY'_r(Y_rY'_r)^{-1}-(A,b)|=0$.
We complete the proof. \hfill$\square$

We can find that the diminishing stepsize condition $\lim_{k\rightarrow\infty}\alpha_k=0$ given in Theorem \ref{th2}
naturally holds under the condition $\sum^\infty_{k=0}\alpha_k^2<\infty$, which is sufficient by Theorem \ref{optimal}
and somehow necessary for the optimal convergence of subgradient almorithms.
The result in theorem \ref{th2} implies that the synchronous optimization algorithm (\ref{eqq1}) is not
privacy preserving in the sense that the malicious agent can discover the adjacency matrix and then other
agents' subgradients asymptotically by choosing an appropriate data sequence transmitted to other regular agents.
In fact, according to the proof of Theorem \ref{th2}, $\lim_{r\rightarrow\infty}A_r=A$,
$\lim_{r\rightarrow\infty}b_r=b$, where $(A_r,b_r)$ is the matrix pair such that $Z_rY'_r(Y_rY'_r)^{-1}:=(A_r,b_r)$.
Then
we can find that regular agents' subgradients at any time $k$ can be obtained
approximately by
$$
\frac{A_rx(k)+b_ru(k)-x(k+1)}{\alpha_k}
$$
with sufficiently large $r$. Under the assumption that the malicious agent knows the adjacency matrix,
Yan et al. showed that the malicious agent can discover other regular agents'
subgradients if and only if all other regular
agents are the malicious agent's neighbors \cite{yan}.
This is consistent with our result.

\section{A Distributed Subgradient Projection Asynchronous Optimization Algorithm}

In last section, we showed that when the malicious agent can
observe all other regular agents' estimates, for almost all adjacency matrices except a zero Lebesgue
measure weight set, DSSOA (\ref{eqq1}) is not privacy preserving in the sense that
the adjacency matrix and then regular agents' subgradients can be discovered by the malicious agent asymptotically.
In this section, we will propose a new privacy preserving distributed subgradient projection asynchronous optimization algorithm
and strictly establish its optimal convergence.

The main design idea of the newly proposed privacy preserving distributed subgradient projection asynchronous optimization algorithm is that agents
make their optimization updates asynchronously and we artificially introduce a projection set in the estimate
iterations.

\emph{\textbf{Distributed Subgradient Projection Asynchronous Optimization Algorithm:}}
\begin{align}\label{pp}
x_i(k+1)=\left\{
  \begin{array}{lll}
    P_X\big(\sum_{j\in \mathcal{N}_i}a_{ij}x_j(k)-\frac{1}{r}d_i(k)\big), & \\
    \qquad\qquad\hbox{if $k=\kappa_i(r)$ for some $r$;} \\
    P_X\big(\sum_{j\in \mathcal{N}_i}a_{ij}x_j(k)\big),\hbox{otherwise,} &
  \end{array}
\right.k\geq0,
\end{align}
where $P_X$ denotes the convex projection operator,
$\kappa_i(r)$ is the time when agent $i$ makes its $r$-th time optimization update,
$\{\kappa_i(r)\}_{r\geq1}$ is referred to as agent $i$'s optimization update time sequence, which is deterministic
and only known by agent $i$.
Note that in algorithm (\ref{pp}) agents make their optimization updates
asynchronously and the optimization update time sequences are different for different agents. Here we artificially introduce a bounded convex projection set $X$,
which is known by all agents and contains all the optimal points of $\min\sum^n_{i=1}f_i$.
We can find that both the optimal solutions of $\min\sum^n_{i=1}f_i$
and $\min_X\sum^n_{i=1}f_i$ are identical.

In asynchronous algorithm (\ref{pp}),
after taking a weighted average of the estimates received from its neighbors,
each agent will take a subgradient optimization step and a projection onto set $X$ to generate the estimate at the next step if  the current time is this agent's optimization update time,
and will just take the projection of the weighted average onto set $X$ as the estimate at the next step otherwise.
Here agent $i$'s optimization update time sequence $\{\kappa_i(r)\}_{r\geq1}$
can be given by agent $i$ in advance before algorithm execution,
and can also be correspondingly defined depending on whether
agents make their optimization updates at $k\geq0$.

\begin{rem}
In algorithm (\ref{pp}), after taking a weighted average of the estimates received from
their neighbors and before generating the estimates at the next step,
agents choose to make an optimization update or not. That is, agents make their optimization updates
just at some times. In fact, this intermittent optimization update mechanism have appeared in the literature,
for instance, \cite{lou3,shig,jak,sri,ned}. In \cite{jak,sri,ned} agents choose
to make their optimization updates or not randomly, and the stepsize
is random and taken as the inverse (or the power of the inverse) of the number of all optimization update times
up to the current time.
Different from them, the stepsize is deterministic in our algorithm.
In fact, we can find that
the randomized unconstrained optimization algorithms are not privacy preserving in some sense since based on the results in last section,
the malicious node can discover other agents' stepsizes and then the subgradients
with a positive probability if the malicious agent takes the full knowledge of the adjacency matrix
and can observe all other agents' estimates.
\end{rem}

\begin{rem}
The stepsize choice is extremely important for the optimal convergence of distributed subgradient algorithms. In fact,
Theorems 4.2 and 4.4 in \cite{lou} show that for a network graph
with doubly stochastic adjacency matrix, the optimal convergence of the sum
objective function may be not achieved if the stepsizes are different for different agents.
In fact, both the left eigenvector of adjacency matrix and the stepsizes determine
the weighted sum objective function to be minimized. However, the results in last section
 show that the identical stepsize design and simultaneous optimization update mechanism make
synchronous algorithm (\ref{eqq1}) not privacy preserving.
In the new asynchronous algorithm (\ref{pp}), the stepsize is taken as
the inverse of the times that agents make their optimization updates up to the current time,
similar to that in \cite{jak,sri,ned}. The following
result shows that the optimal convergence can still be guaranteed provided that for each agent,
the number of its optimization update times is the same over different time intervals with the same length.
\end{rem}

\begin{rem}
In cryptograph-based methods \cite{glo}, agents need to encrypt the
estimates needed to be shared with their neighbors and decrypt the received estimates so that agents' privacy
cannot be disclosed. In differential privacy
methods \cite{huang,li,hans,man,dwo}, agents need to add random noises
on the estimates needed to transmitted to protect agents' privacy.
Different from them, in our algorithm (\ref{pp}), the estimates can be transmitted
directly between neighbors without any additional
technique to disguise agents' estimates. It reveals that only the asynchronous optimization
update mechanism can ensure that the proposed algorithm is privacy preserving.
\end{rem}

\begin{rem}
In algorithm (\ref{pp}), for the unconstrained optimization problem
$\min\sum^n_{i=1}f_i$, we artificially introduce a projection set from the viewpoint of privacy preservation.
We can find that algorithm (\ref{pp}) also works for the constrained optimization problem $\min_{K}\sum^n_{i=1}f_i$,
where $X$ can be taken as a subset that contains all the optimal solutions of $\min_{K}\sum^n_{i=1}f_i$.
In fact, the authors in \cite{yan} have shown that in presence of projection set, the malicious agent cannot discover the subgradients for any network graph in their model.
\end{rem}

\subsection{Privacy Preserving Properties}

Before establishing the optimal convergence of algorithm (\ref{pp}),
in this subsection we first roughly illustrate that algorithm (\ref{pp}) is privacy preserving
from the two aspects of projection set $X$ and asynchronous optimization update mechanism.

First, when agents' ``estimates"  $\sum_{j\in \mathcal{N}_i}a_{ij}x_j(k)-\frac{1}{r}d_i(k)$ locate
outside the projection set $X$, from the property of convex projection operator
$$P_X(z)=P_X\big(P_X(z)+\lambda(z-P_X(z))\big)\footnote{This property of convex projection operator
follows from the fact that $w=P_X(z)$ if and only if $(z-w)'(y-w)\leq0$ for any $y\in X$.
This fact can be shown directly from the definition of convex projection.},\forall z\not\in X,\;\lambda\geq0,$$
we know that the malicious agent cannot infer
other agents' subgradients at time $k$ based on its received estimates even though the malicious agent knows the adjacency matrix, while
when they locate inside set $X$, algorithm (\ref{pp}) evolves
in the form:
$$
x_i(k+1)=\left\{
  \begin{array}{lll}
    \sum_{j\in \mathcal{N}_i}a_{ij}x_j(k)-\frac{1}{r}d_i(k), & \\
    \qquad\qquad\hbox{if $k=\kappa_i(r)$ for some $r$;}&\\
    \sum_{j\in \mathcal{N}_i}a_{ij}x_j(k),\;\;\hbox{otherwise.}&
  \end{array}
\right.
$$
This also reveals that the malicious agent cannot discover
other agents' subgradients at time $k$ based on the following reasons.
On one hand, even if the adjacency matrix $(A,b)$ has been obtained by the malicious agent
and the malicious agent can observe all other regular agents' estimates, but note that
since the malicious agent does not know whether regular agents $i,i\neq n$ make their optimization updates at time $k$, so in this asynchronous algorithm, knowing $x_i(k+1)-\sum_{j\in \mathcal{N}_i}a_{ij}x_j(k)$
cannot help the malicious agent discover the subgradients;
on the other hand, even if the malicious agent also knows that agent $i$ makes its optimization update at time $k$,
which helps the malicious agent discover $\frac{1}{r}d_i(k)$ from
$x_i(k+1)-\sum_{j\in \mathcal{N}_i}a_{ij}x_j(k)$, but this malicious agent still cannot
discover the subgradient since it does not know the
stepsize $1/r$ considering that the optimization update time sequences are different for different agents and each agent only knows its own update time sequence.

As a sum, we can roughly conclude that when agents are far from the
projection set $X$, both the projection set and the asynchronous optimization
update mechanism can effectively protect agents' privacy and when agents
are close to the desired optimal solution $x^*\in \arg\min\sum^n_{i=1}f_i$ (the optimal convergence will
be proven in the following Theorem), it is the asynchronous optimization
update mechanism that mainly protects agents' privacy.

\subsection{Optimal Convergence}

In this subsection, we will establish the optimal convergence of the newly proposed
asynchronous projection optimization algorithm (\ref{pp}).
We next make an assumption on agents' optimization update time sequences $\{\kappa_i(r)\}_{r\geq1}$, $i=1,...,n$.

\noindent {\bf Assumption 4}: There exists an integer $T>0$
such that for each agent $i$, $1\leq t_i(r_1)=t_i(r_2)<\infty$, $\forall r_1,r_2$, where
$$
t_i(r)=\Big|\big\{s\big|rT\leq\kappa_i(s)<(r+1)T\big\}\Big|
$$
denotes the times of agent $i$'s optimization updates on interval $[rT,(r+1)T)$.

Assumption 4 requires that each agent makes its own optimization update
with a constant number of times within any time interval with some fixed common length.
Note that the numbers of optimization updates within the time interval with the fixed length
may be different for different agents. We can see that Assumption 4 holds if
each agent makes its optimization update in a periodic way, no matter whether the periods
of agents' optimization updates are the same.

We now establish the optimal convergence of algorithm (\ref{pp}).
\begin{them}(\textbf{Optimal Convergence})
Consider distributed subgradient projection asynchronous optimization algorithm (\ref{pp}) with Assumptions
1, 2 and 4.
Then the network will achieve an optimal consensus, i.e., there exists $x^*\in \arg\min\sum^n_{i=1}f_i$ such that $\lim_{k\rightarrow\infty}x_i(k)=x^*$, $i=1,...,n$.
\end{them}
\emph{Proof.}
First it follows from $x_i(k)\in X$
that $\sum_{j\in \mathcal{N}_i}a_{ij}x_j(k)\in X$ by Assumption 2 and the convexity of $X$. Then for $k\geq1$,
algorithm (\ref{pp}) can be re-written as
\begin{align}\label{pp2}
x_i(k+1)=\left\{
  \begin{array}{lll}
    \sum_{j\in \mathcal{N}_i}a_{ij}x_j(k)+\omega_i(k), & \\
    \qquad\qquad\hbox{if $k=\kappa_i(r)$ for some $r$;} \\
    \sum_{j\in \mathcal{N}_i}a_{ij}x_j(k),\;\;\hbox{otherwise,} &
  \end{array}
\right.
\end{align}
where
$$
\omega_i(k)=P_X\Big(\sum_{j\in \mathcal{N}_i}a_{ij}x_j(k)-\frac{1}{r}d_i(k)\Big)-\sum_{j\in \mathcal{N}_i}a_{ij}x_j(k).
$$

Here we still use $L$ to denote the upper bound of subgradients
of objective functions,
$$L:=\sup_{q\in\bigcup_{x\in X,i}\partial f_i(x)}|q|,$$
which is a finite number due to the boundedness of $X$ and the convexity of $f_i$.
This implies that Assumption 3 holds.
Therefore,
$$|\omega_i(k)|\leq \frac{1}{r}|d_i(k)|\leq\frac{1}{r}L$$
and then it follows from Assumption 4 that $\lim_{k\rightarrow\infty}|\omega_i(k)|=0$,
where we use the property of convex projection operator
$|P_X(y)-z|\leq|y-z|$ for any $y\in \mathbb{R}^m$ and $z\in X$\footnote{This property of convex projection operation
comes from Lemma 1 (b) in \cite{nedic2}.} and the fact that $\sum_{j\in \mathcal{N}_i}a_{ij}x_j(k)\in X$.
As a result, algorithm (\ref{pp}) will achieve a consensus, i.e.,
$\lim_{k\rightarrow\infty}h(k)=0$
 by Theorem 1 in \cite{wang}, where
 $$h(k)=\max_{i,j}|x_i(k)-x_j(k)|.$$

Let $x^*\in\arg\min_X\sum^n_{i=1}f_i$. Then by applying the similar analysis for distributed subgradient algorithms in \cite{nedic1,nedic2,lou}, we have that when $k=\kappa_i(r)$ for some $r$,
\begin{align}\label{do3}
&|x_i(k+1)-x^*|^2\nonumber\\
&=\Big|P_X\Big(\sum_{j\in \mathcal{N}_i}a_{ij}x_j(k)-\frac{1}{r}d_i(k)\Big)-x^*\Big|^2\nonumber\\
&\leq\Big|\sum_{j\in \mathcal{N}_i}a_{ij}x_j(k)-\frac{1}{r} d_i(k)-x^*\Big|^2\nonumber\\
&\leq\Big|\sum_{j\in \mathcal{N}_i}a_{ij}x_j(k)-x^*\Big|^2+\frac{|d_i(k)|^2}{r^2}
-\frac{2}{r}(x_i(k)-x^*)'d_i(k)+\frac{2L}{r}\Big|x_i(k)-\sum_{j\in \mathcal{N}_i}a_{ij}x_j(k)\Big|\nonumber\\
&\leq\sum_{j\in \mathcal{N}_i}a_{ij}\big|x_j(k)-x^*\big|^2+\frac{L^2}{r^2}
-\frac{2}{r}\big(f_i(x_i(k))-f_i(x^*)\big)+\frac{2L}{r}\Big|x_i(k)-\sum_{j\in \mathcal{N}_i}a_{ij}x_j(k)\Big|\nonumber\\
&\leq\sum_{j\in \mathcal{N}_i}a_{ij}\big|x_j(k)-x^*\big|^2+\frac{L^2}{r^2}
-\frac{2}{r}\big(f_i(\bar x(k))-f_i(x^*)\big)\nonumber\\
&\qquad+\frac{2L}{r}\Big(\big|x_i(k)-\sum_{j\in \mathcal{N}_i}a_{ij}x_j(k)\big|+|x_i(k)-\bar x(k)|\Big)\nonumber\\
&\leq\sum_{j\in \mathcal{N}_i}a_{ij}\big|x_j(k)-x^*\big|^2+\frac{L^2}{r^2}
-\frac{2}{r}\big(f_i(\bar x(k))-f_i(x^*)\big)+\frac{4L}{r}h(k),\nonumber
\end{align}
where $\bar x(k)=\frac{1}{n}\sum^n_{i=1}x_i(k)$ denotes the average of agents' estimates at time $k$.
Moreover, when $k\not=\kappa_i(r)$ for any $r$, we have
\begin{align}
|x_i(k+1)-x^*|^2&=\Big|\sum_{j\in \mathcal{N}_i}a_{ij}x_j(k)-x^*\Big|^2\leq\sum_{j\in \mathcal{N}_i}a_{ij}|x_j(k)-x^*|^2.\nonumber
\end{align}
Summarizing the above two cases, we have
\begin{align}
&|x_i(k+1)-x^*|^2\leq\sum_{j\in \mathcal{N}_i}a_{ij}\big|x_j(k)-x^*\big|^2+\chi_{i,k}\Big(\frac{L^2}{r^2}
-\frac{2}{r}\big(f_i(\bar x(k))-f_i(x^*))+\frac{4L}{r}h(k)\Big),\nonumber
\end{align}
where
$$
\chi_{i,k}=\left\{
  \begin{array}{ll}
    1, & \hbox{if $k=\kappa_i(r)$ for some $r$;} \\
    0, &\hbox{otherwise}
  \end{array}
\right.
$$
Taking the sum of the above inequality over $i=1,...,n$, by the double stochasticity in Assumption 2 we have
\begin{align}
&\sum^n_{i=1}|x_i(k+1)-x^*|^2\leq\sum^n_{i=1}|x_i(k)-x^*|^2+\sum^n_{i=1}\chi_{i,k}\Big(\frac{L^2}{r^2}
-\frac{2}{r}\big(f_i(\bar x(k))-f_i(x^*)\big)+\frac{4L}{r}h(k)\Big)\nonumber
\end{align}
As a result,
\begin{align}
&\sum^n_{i=1}|x_i((s+1)T)-x^*|^2
\leq\sum^n_{i=1}|x_i(sT)-x^*|^2 +\sum^{(s+1)T-1}_{k=sT}\sum^n_{i=1}\chi_{i,k}\frac{L^2}{r^2}\nonumber\\
&\qquad
-\sum^{(s+1)T-1}_{k=sT}\sum^n_{i=1}\chi_{i,k}\frac{2}{r}\big(f_i(\bar x(k))-f_i(x^*)\big)+\sum^{(s+1)T-1}_{k=sT}\sum^n_{i=1}\chi_{i,k}\frac{4L}{r}h(k)\nonumber\\
&:=\sum^n_{i=1}|x_i(sT)-x^*|^2+\mu_1(s)+\mu_2(s)+\mu_3(s)
\end{align}

We next estimate the sum of the second, third and fourth term in (12)
over $s\geq1$.
Then by the condition $t_i(s_1)=t_i(s_2)$ for all $s_1,s_2$ in Assumption 4, we have
\begin{align}\label{add2}
\sum^\infty_{s=1}\mu_1(s)
&=\sum^\infty_{s=1}\sum^n_{i=1}\sum^{t_i(s)}_{r=1}\frac{L^2}{\big(t_i(0)+\cdots+t_i(s-1)+r\big)^2}\nonumber\\
&\leq\sum^\infty_{s=1}\sum^n_{i=1}\frac{t_i(s)L^2}{\big(t_i(0)+\cdots+t_i(s-1)\big)^2}\nonumber\\
&=\sum^\infty_{s=1}\frac{1}{s^2}\sum^n_{i=1}\frac{L^2}{t_i(0)}<\infty
\end{align}

We also have
\begin{align}\label{add1}
\mu_2(s)
&=-\sum^n_{i=1}\sum^{t_i(s)}_{r=1}\frac{2}{t_i(0)+\cdots+t_i(s-1)+r}\big(f_i(\bar x(\kappa_i(t_i(0)+\cdots+t_i(s-1)+r)))-f_i(x^*)\big)\nonumber\\
&=-\sum^n_{i=1}\sum^{t_i(s)}_{r=1}\frac{2}{t_i(0)+\cdots+t_i(s-1)}(f_i(\bar x(sT))-f_i(x^*))\nonumber\\
&\;\;\;\;-\sum^n_{i=1}\sum^{t_i(s)}_{r=1}\frac{2}{t_i(0)+\cdots+t_i(s-1)}(f_i(\bar x(\kappa_i(t_i(0)+\cdots+t_i(s-1)+r)))-f_i(\bar x(sT)))\nonumber\\
&\;\;\;\;-\sum^n_{i=1}\sum^{t_i(s)}_{r=1}
\Big(\frac{2}{t_i(0)+\cdots+t_i(s-1)+r}-\frac{2}{t_i(0)+\cdots+t_i(s-1)}\Big)\nonumber\\
&\qquad\times(f_i(\bar x(\kappa_i(t_i(0)+\cdots+t_i(s-1)+r)))-f_i(x^*))\nonumber\\
&\leq-\frac{2}{s}\sum^n_{i=1}(f_i(\bar x(sT))-f_i(x^*))\nonumber\\
&\;\;\;\;+\sum^n_{i=1}\frac{2Lt_i(s)}{t_i(0)+\cdots+t_i(s-1)}\max_{sT\leq r<(s+1)T}|\bar x(r)-\bar x(sT)|\nonumber\\
&\;\;\;\;+\sum^n_{i=1}\sum^{t_i(s)}_{r=1}\frac{2Lr}{(t_i(0)+\cdots+t_i(s-1))^2}\max_{sT\leq r<(s+1)T}|\bar x(r)-x^*|.
\end{align}

Taking the average of the two sides of (\ref{pp2}), by Assumption 2 we have
$$\bar x(k+1)=\bar x(k)+\frac{1}{n}\sum^n_{i=1}\chi_{i,k}\omega_i(k)$$
and then
\begin{align}
\max_{sT\leq r<(s+1)T}|\bar x(r)-\bar x(sT)|
&\leq\sum^{(s+1)T-2}_{k=sT}\frac{1}{n}\sum^n_{i=1}\chi_{i,k}|\omega_i(k)|\nonumber\\
&\leq\frac{1}{n}\sum^n_{i=1}\sum^{t_i(s)}_{r=1}\frac{L}{t_i(0)+\cdots+t_i(s-1)+r}\nonumber\\
&\leq\frac{L}{s}.\nonumber
\end{align}
This implies
\begin{align}\label{add3}
&\sum^\infty_{s=1}\sum^n_{i=1}\frac{2Lt_i(s)}{t_i(0)+\cdots+t_i(s-1)}\max_{sT\leq r<(s+1)T}|\bar x(r)-\bar x(sT)|\nonumber\\
&\leq\sum^\infty_{s=1}\frac{2L^2n}{s^2}<\infty
\end{align}
Moreover, we also have
\begin{align}\label{add5}
&\sum^\infty_{s=1}\sum^n_{i=1}\sum^{t_i(s)}_{r=1}\frac{2Lr}{(t_i(0)+\cdots+t_i(s-1))^2}\max_{sT\leq r<(s+1)T}|\bar x(r)-x^*|\nonumber\\
&\leq\sum^\infty_{s=1}\sum^n_{i=1}\sum^{t_i(s)}_{r=1}\frac{2L\zeta r}{(t_i(0)+\cdots+t_i(s-1))^2}\nonumber\\
&\leq\sum^\infty_{s=1}\frac{1}{s^2}\sum^n_{i=1}\frac{2L\zeta}{(t_i(0))^2}\frac{(1+t_i(0))t_i(0)}{2}\nonumber\\
&<\infty,
\end{align}
where $\zeta=\sup_s\max_{sT\leq r<(s+1)T}|\bar x(r)-x^*|<\infty$ by the boundedness of $X$
and the fact $\bar x(r)\in X$.
Combining with (\ref{add1}), (\ref{add3}) and (\ref{add5})
together, we have
\begin{align}\label{add6}
\sum^\infty_{s=1}\big(\mu_2(s)+\frac{2}{s}\sum^n_{i=1}(f_i(\bar x(sT))-f_i(x^*))\big)<\infty.
\end{align}

By the similar arguments given in the proof of Lemma 4.3 in \cite{lou}, we can also show that
\begin{align}\label{add7}
\sum^\infty_{s=1}\mu_3(s)
&\leq\sum^\infty_{s=1}\sum^n_{i=1}\sum^{t_i(s)}_{r=1}\frac{4L}{t_i(0)+\cdots+t_i(s-1)+r}\max_{sT\leq k<(s+1)T}h(k)\nonumber\\
&\leq\sum^\infty_{s=1}\sum^n_{i=1}\frac{4Lt_i(s)}{t_i(0)+\cdots+t_i(s-1)}\max_{sT\leq k<(s+1)T}h(k)\nonumber\\
&=\sum^\infty_{s=1}\frac{4Ln}{s}
\max_{sT\leq k<(s+1)T}h(k)<\infty.
\end{align}

By (11), (\ref{add2}), (\ref{add6}) and (\ref{add7}), we have
\begin{align}
\sum^n_{i=1}|x_i((s+1)T)-x^*|^2&\leq\sum^n_{i=1}|x_i(sT)-x^*|^2-\frac{2}{s}\sum^n_{i=1}(f_i(\bar x(sT))-f_i(x^*))\nonumber\\
&+\mu_1(s)+\mu_2(s)+\frac{2}{s}\sum^n_{i=1}(f_i(\bar x(sT))-f_i(x^*))+\mu_3(s)\nonumber
\end{align}
with
$$\sum^\infty_{s=1}\big(\mu_1(s)+\mu_2(s)+\frac{2}{s}\sum^n_{i=1}(f_i(\bar x(sT))-f_i(x^*))+\mu_3(s)\big)<\infty.$$
Then we conclude that the limit $\lim_{s\rightarrow\infty}|x_i(sT)-x^*|^2$ exists
and $\sum^\infty_{s=1}\frac{2}{s}\sum^n_{i=1}(f_i(\bar x(sT))-f_i(x^*))<\infty$.
Since $\sum^\infty_{s=1}\frac{2}{s}=\infty$,
$$\liminf_{s\rightarrow\infty}\sum^n_{i=1}(f_i(\bar x(sT))-f_i(x^*))=0.$$

Let $\{\bar x(s^rT)\}_{r\geq0}$ be a subsequence of $\{\bar x(sT)\}_{s\geq 0}$ such that
 $\lim_{r\rightarrow\infty}\sum^n_{i=1}(f_i(\bar x(s^rT))-f_i(x^*))=0$.
From the boundedness of $\{\bar x(s^rT)\}$, we know that there exists a subsequence $\{\bar x(s^{r_k}T)\}_{k\geq0}$
of $\{\bar x(s^rT)\}_{r\geq0}$ such that the limit $\lim_{k\rightarrow\infty}\bar x(s^{r_k}T)=\hat x$ exists. Therefore, it follows from the continuity of $f_i$ and the closedness of $X$ that $\hat x\in\arg\min_X\sum^n_{i=1}f_i$.
By replacing $x^*$ with $\hat x$, we can also similarly show that
the limit $\lim_{s\rightarrow\infty}\sum^n_{i=1}|x_i(sT)-\hat x|^2$ exists.
This combines with $\lim_{k\rightarrow\infty}\bar x(s^{r_k}T)=\hat x$ and
what we have shown that the consensus is achieved imply
that $\lim_{s\rightarrow\infty}\sum^n_{i=1}|x_i(sT)-\hat x|^2=0$.

We complete the proof. \hfill$\square$

\section{Conclusion}

In this paper, we considered the privacy preserving features
of distributed subgradient optimization algorithms. We first
show that an existing distributed subgradient synchronous optimization
algorithm is not privacy preserving in the sense that
the malicious agent can learn
other agents' subgradients asymptotically for almost all adjacency matrices except a zero Lebesgue measure weight set.
We also proposed a new distributed subgradient projection asynchronous optimizaiton
algorithm, in which agents make their own optimization updates asynchronously and each agent
only knows its own optimization update time sequence. The artificially introduced
convex projection set and the asynchronous optimization mechanism can effectively
protect agents' private information. Moreover, we also shown the optimal convergence
of the newly proposed asynchronous algorithm. Other interesting problems,
including investigating privacy preserving properties of other distributed
optimization algorithms such as subgradient random algorithms \cite{jak,sri,ned},
dual averaging algorithm \cite{duc} and ADMM \cite{boyd,shiw}, and developing other
privacy preserving algorithms using the proposed privacy preservation techniques in this paper, are still under investigation.

\vspace{5mm}
\noindent \textsc{Youcheng Lou and Shouyang Wang} \\
\noindent Academy of Mathematics and Systems Science\\
\noindent Chinese Academy of Sciences \\
\noindent Beijing 100190, China \\
\noindent Email: {\tt\small louyoucheng@amss.ac.cn, sywang@amss.ac.cn} \\

\noindent \textsc{Lean Yu} \\
\noindent School of Economics and Management\\
\noindent Beijing University of Chemical Technology \\
\noindent Beijing 100029, China \\
\noindent Email: {\tt\small yulean@amss.ac.cn} \\
\end{document}